\documentclass{aa}
\usepackage{times,graphicx,epsfig}
\newcommand{\bq}{\begin{equation}}
\newcommand{\eq}{\end{equation}}
\newcommand{\bqn}{\begin{eqnarray}}
\newcommand{\eqn}{\end{eqnarray}}
\newcommand{\dd}{\mbox{\rm d}}
\newcommand{\msun}{\rm{M}_\mathrm{\rm \sun}}

\begin{document}
\title
{The local star formation history of the thin disc derived from kinematic data}
\author{Andreas Just\inst{1}\and
        Hartmut Jahrei\ss\inst{1}
}

\offprints{A. Just}
\mail{just@ari.uni-heidelberg.de}
\institute{Astronomisches Rechen-Institut at ZAH, M\"onchhofstra{\ss}e 12-14, 69120
Heidelberg, Germany
}

\date{Printed: \today}

\abstract
{We present an evolutionary disc model for the thin disc in the solar cylinder
 based on a continuous  star formation history (SFR) 
 and a continuous dynamical heating (AVR) of
the stellar subpopulations.} 
{We determine the star formation history of the thin disc in the solar
vicinity. The self-consistent model of the vertical structure allows predictions
of the density, age, metallicity and velocity distribution of main sequence
stars as a function of height above the midplane.} 
{The vertical distribution of the stellar
subpopulations are calculated self-consistently in dynamical equilibrium. The
SFR and AVR of the stellar subpopulations are determined by
fitting the velocity distribution functions of main sequence stars.} 
{We find a vertical disc model for the thin disc including the gas and dark
matter component, which is consistent with the
local kinematics of main sequence stars and fulfils the known constraints on the
surface densities and mass ratios. The SFR
shows a maximum 10\,Gyr ago declining by a factor of 10 until present time.
The velocity dispersion of the stellar
subpopulations increase with age according to a power law with index 0.375. 
Using the new scale heights leads to a best fit IMF with power-law indices of 
1.5 below and 4.0 above 1.6$\msun$, which has no kink around 1$\msun$.
Including a thick disc component results in slight variations of the
thin disc properties, but has a negligible influence on the SFR.
A variety of predictions are made concerning the number density, age 
and metallicity distributions of stellar subpopulations as a function of $z$
above the galactic plane.}
{ The combination of kinematic data from Hipparcos and the finite lifetimes of
main sequence stars allows strong constraints on the structure and history of
the local disc. 
A constant SFR can be ruled out.}

\titlerunning{Local SFR of the thin disc}

\authorrunning{A. Just, H. Jahrei\ss}
\maketitle
\section{Introduction\label{introd}}

The star formation history $SFR(t)$ of the Milky Way disc is still not very well
determined. The main reason for that is the lack of good age estimators with
corresponding unbiased stellar samples. Especially samples selected by colour
cuts or by a magnitude limit are biased with respect to the age distribution,
because there is an age-metallicity relation due to the chemical enrichment of 
the disc. Therefore not even the famous Geneva-Copenhagen sample of 14.000 F and G stars
(Nordstroem et al. \cite{nor04}) with well determined stellar properties
including individual age estimates can be used to derive the star formation
history directly by star counts. 
The most complete disc model based on star counts is that of 
Robin et al. (\cite{rob03}). They used a series of stellar
subpopulations of different ages including the AVR and the chemical enrichment
but fixing the $SFR$ to be constant. As a result, scale heights and surface
densities of the stellar subpopulations are systematically smaller 
than in our model.
Rocha-Pinto et al. (\cite{roc00}) used chromospheric age determinations of late
type dwarfs. They apply stellar evolution and scale height corrections. The
main result is the determination of enhanced star formation episodes over the
lifetime of the disc. They exclude a constant $SFR$ from chemical evolution
models. In R. \& C. de la Fuente Marcos (\cite{fue04}) the star formation history
for open star clusters were determined. 
They found at least five episodes of enhanced star formation in the last 2\,Gyr.
Since star cluster contain only a small
percentage of all disc stars, an extrapolation to the total (smoothed) $SFR$ is
not possible. 

In recent years the Hipparcos stars with precise parallaxes and proper motions
were used to determine the $SFR$ with different methods.
Hernandez et al. (\cite{her00}) determined the $SFR$ over
the last 3\,Gyr using isochrone ages. They found a series of star formation episodes on top of  an
underlying smooth $SFR$. This result is complementary to our model, which gives
the slow evolution of the smoothed $SFR$. 
In Binney et al. (\cite{bin00}) and in Cignoni et al. (\cite{cig06}) the scale
height variation of main sequence stars were not taken into account. Therefore
these models derive the local age distribution of K and M dwarfs (with
lifetimes larger than the age of the disc) instead of the $SFR$. Binney et al.
 (\cite{bin00})
determined an age of the thin disc of 11.2\,Gyr (consistent with the 12\,Gyr we
are using) and both papers
found an approximately constant age distribution, which is fully consistent with
our model.

We present the first disc model, which determines the $SFR$, the $AVR$ and scale
heights consistently.
We use the Hipparcos stars and at the faint end the Catalogue of Nearby Stars
(CNS4) to construct an evolutionary disc model including the kinematic
information. We select the main sequence stars and divide these into a series of
volume complete subsamples from B to K type. In a kinematic model we derive
the vertical velocity distribution functions $f(|W|)$ of each subsample, which
depends mainly on the star formation history and the dynamical evolution of the
disc described by the dynamical heating $\sigma_\mathrm{W}(t)$ 
(the age velocity dispersion relation AVR). The derived $SFR$ depends on the
normalized velocity distribution functions only. Therefore the model is
essentially independent of the adopted $IMF$. In a final step we use the
disc model to derive the local $IMF$ from the volume complete subsamples
applying consistently the lifetime and the scale height corrections.

\section{Disc Model \label{model}}

In this section we describe the construction of the disc model and the
determination of the properties of the stellar subsamples. For the
self-consistent determination of the gravitational potential we take 
into account the stellar
component, the gas component and the dark matter halo.

We use a thin, self-gravitating disc composed of a sequence of stellar
subpopulations according to the $SFR$ and the AVR. Additionally a cold gas
component and a Dark Matter halo are included for the gravitational forces. We
take into account finite lifetimes of the stars and mass loss due to stellar
evolution. The stellar remnants stay in their subpopulation, the expelled gas
(stellar winds, PNs, and SNs) is mixed implicitly into the gas component.
Since the lifetimes and mass loss depend on metallicity, the metal
enrichment with time is included. A standard  $IMF$ is used for the determination of
the stellar mass fraction of the subpopulations with age.

From large sets of template functions for the $SFR$ and AVR we search for the
best fitting pair of input functions. The total gravitational potential and 
the density profile of each age-bin are determined self-consistently assuming
isothermal distribution functions with velocity dispersion according to the
AVR. The velocity distribution functions $f(|W|)$ for main sequence stars are
calculated by a superposition of the Gaussians weighted by the local density
contribution up to the lifetime of the stars. These distribution functions are
simultaneously compared to the observed
$f(|W|)$ in magnitude bins along the main sequence to judge the quality of the
model.

\subsection{Self-gravitating disc \label{grav}}

The backbone of the disc is a self-gravitating vertical disc profile including 
the gas component in the thin disc approximation. In this approximation the
Poisson-Equation is one-dimensional
and in the case of a purely self-gravitating thin disc
(i.e. no external potential) the Poisson equation can be
integrated leading to
\bq
\left(\frac{\dd\Phi}{\dd z}\right)^2
        =8\pi G \int_0^{\Phi}\rho(\Phi')\dd\Phi' \quad.\label{eqkz}
\eq
Therefore we model all gravitational components by a thin disc approximation.
We include in the total
potential $\Phi$, the stellar component $\Phi_\mathrm{s}$, 
the gas component $\Phi_\mathrm{g}$, 
and the dark matter halo contribution $\Phi_\mathrm{h}$
\bq
\Phi(z)=\Phi_\mathrm{s}(z)+\Phi_\mathrm{g}(z)+\Phi_\mathrm{h}(z)\quad.\label{eqpot}
\eq
In order to obtain the force of a
spherical halo correctly in the thin disc approximation,
 we use a special approximation
(see subsection \ref{halo}).
Since we construct the disc in dynamical 
equilibrium, the density of the sub-components are given as a function 
of the total potential $\rho_\mathrm{j}(\Phi)$.
The vertical distribution is given by the implicit function $z(\Phi)$ 
via direct integration
\bq
z(\Phi)=\int_0^{\Phi}\dd\Phi'
        \left[8\pi G \int_0^{\Phi'}\rho(\Phi'')\dd\Phi''\right]^{-1/2}
         \quad.\label{eqz}
\eq

 The relative
contribution of the stellar, the gaseous, and the DM-component to the surface
density (up to $|z|=z_\mathrm{max}$) are given  by the input parameters
$Q_\mathrm{s},Q_\mathrm{g},Q_\mathrm{h}$, which have to be iterated to fit all
observational constraints.
The disc model has two global scaling parameters to convert the normalized model
to physical quantities. We fix the local stellar density to
$\rho_\mathrm{s0}=0.039\,\msun\,pc^{-3}$,
which is the best observed quantity for the local disc model 
(Jahrei{\ss} et al. \cite{jah97}). The second scaling parameter is the
exponential scale height $z_\mathrm{s}$, which fixes via the shape correction the
half-thickness of the stellar disc $h_\mathrm{eff}$, with the mass
fractions $Q_\mathrm{s},Q_\mathrm{g},Q_\mathrm{h}$ the surface densities, 
and via
the maximum velocity dispersion $\sigma_\mathrm{e}$ the scaling of the velocity distribution
functions.

\subsection{Stellar disc \label{stars}}

The stellar component is composed by a sequence of isothermal subpopulations
characterized by the IMF, the chemical enrichment $\mathrm{[Fe/H]}(t)$, 
the star formation history $SFR(\tau)$, and the dynamical evolution described by
the vertical velocity dispersion $\sigma_\mathrm{W}(t)$. Here 
$\tau=t_\mathrm{a}-t$ is the time and $t$ is the age of the
subpopulation running back in time from the present time $t_\mathrm{a}=12$\,Gyr
(which is the adopted age of the disc). 
We include mass loss due to stellar evolution and retain the
stellar-dynamical mass fraction $g(t)$ (stars + remnants) only. The mass lost
by stellar winds, supernovae and planetary nebulae is mixed implicitly to the
gas component. 

With the Jeans equation the vertical
distribution of each isothermal subpopulation is given by
\bq
\rho_\mathrm{s,j}(z)=\rho_\mathrm{s0,j}
	\exp\left( \frac{-\Phi(z)}{\sigma^2_\mathrm{W,j}}\right)
\quad,
\eq
where $\rho_\mathrm{s,j}$ is actually a 'density rate', the density per age bin,
and $\sigma_\mathrm{W,j}$ the velocity dispersion at age $t_\mathrm{j}$. The
connection to the $SFR$ is given by the integral over $z$
\bq
g(t_\mathrm{j})SFR(\tau_\mathrm{j})=
	\int_\mathrm{-\infty}^{\infty}\rho_\mathrm{s,j}(z)\dd z\quad.
\eq
with time $\tau_\mathrm{j}=t_\mathrm{a}-t_\mathrm{j}$.
The (half-)thickness $h_\mathrm{p}(t_\mathrm{j})$ of the subpopulations is 
defined by the midplane density $\rho_\mathrm{s0,j}$ through
\bq
\rho_\mathrm{s0,j}=\frac{g(t_\mathrm{j})SFR(\tau_\mathrm{j})}
			{2h_\mathrm{p}(t_\mathrm{j})}\quad .
\eq
and can be calculated by
\bq
h_\mathrm{p}(t_\mathrm{j})=\int_0^{\infty}
	\frac{\rho_\mathrm{s,j}(z)}{\rho_\mathrm{s0,j}}\dd z=\int_0^{\infty}
	\exp\left( \frac{-\Phi(z)}{\sigma^2_\mathrm{W,j}}\right)\dd z \quad.
\eq
The total stellar density $\rho_\mathrm{s}(z)$ and 
velocity dispersion $\sigma_\mathrm{s}(z)$ are determined by
\bqn
\rho_\mathrm{s}(z)&=&\int_0^{t_\mathrm{a}}\rho_\mathrm{s,j}(z)\dd t \\
\sigma^2_\mathrm{s}(z)&=&\frac{1}{\rho_\mathrm{s}(z)}\int_0^{t_\mathrm{a}}
	\sigma^2_\mathrm{W,j}\rho_\mathrm{s,j}(z)\dd t \quad.
\eqn
$\rho_\mathrm{s}(z)$ then determines the potential $\Phi_\mathrm{s}(z)$ via
the Poisson equation. The total stellar surface density $\Sigma_\mathrm{s}$ is
connected to the integrated star formation $S_0$ by the effective
stellar-dynamical mass fraction $g_\mathrm{eff}$
\bq
\Sigma_\mathrm{s}=\int\rho_\mathrm{s}(z)\dd z=g_\mathrm{eff}S_0\label{eqSigmas}
\eq
with
\bq
S_0=\int SFR\, \dd t\quad \mbox{and}\quad 
        g_\mathrm{eff} =\frac{\int g(t)SFR(\tau)\dd t}{S_0}
	\quad ,\label{eqS0}
\eq
which includes luminous stars and stellar remnants.
The local stellar density is given by
\bq
\rho_\mathrm{s0}=\frac{g_\mathrm{eff}S_0}{2h_\mathrm{eff}}\label{eqrhos0}
\eq
with thickness $h_\mathrm{eff}$ for all stars. For a general shape of the density
profile  $h_\mathrm{eff}$ is different to $z_\mathrm{s}$, the exponential
scale height well above the midplane. 
The effective exponential scale height $z_\mathrm{s}$ of the stellar disc describing the 
exponential decline above the midplane is 
connected to the maximum velocity dispersion $\sigma_\mathrm{e}$ of the 
subpopulations and the total surface
density $\Sigma_\mathrm{tot}$ via
\bq
z_\mathrm{s}=C_\mathrm{z}z_\mathrm{e}=C_\mathrm{z}\frac{\sigma_\mathrm{e}^2}
			{2\pi G \Sigma_\mathrm{tot}}\quad, \label{eqze}
\eq
where $z_\mathrm{e}$ is the exponential scale height of an isothermal component above a disc
with total surface density $\Sigma_\mathrm{tot}$. The 
shape correction factor $C_\mathrm{z}$ is of order unity and is determined by
the mean exponential scale height in the range
 $z=(2 - 5)\,z_\mathrm{e}$.

\begin{figure}[t]
\epsfig{file=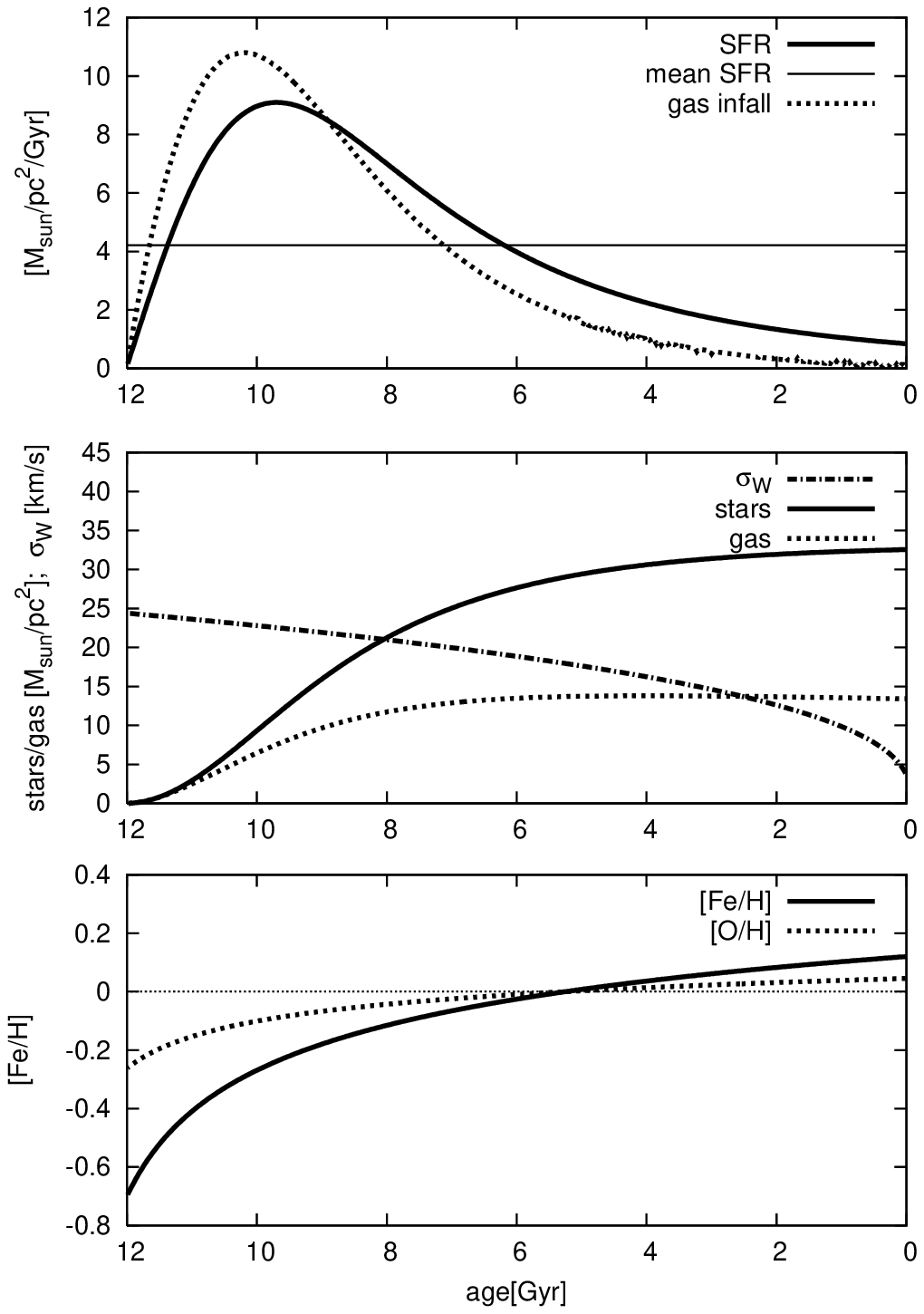,width=8.8cm,angle=0}
\caption[]{
The upper panel shows the $SFR$ for the model and the infall rate of primordial
gas derived from a local chemical evolution model (see Sect.\ \ref{feh}).
The thin line is the mean value of the SFR for comparison.
The next panel gives the velocity dispersion of the stellar subpopulations 
The lower panel shows the adopted metal enrichment $\mathrm{[Fe/H]}(t)$, which leads to
 a local metallicity distribution of late G dwarfs consistent with the
 observations (see Fig.\ \ref{figfeh}).
}
\label{figsfr}
\end{figure}

The metallicity $\mathrm{[Fe/H]}$ affects the lifetimes, luminosities and colours 
of the stars and as a consequence also the mass loss of the
subpopulations. 
In order to account for the influence of the metal enrichment
we include a metal enrichment $\mathrm{[Fe/H]}(t)$ (see Fig.\ \ref{figsfr}), which leads to
 a local metallicity distribution of late G dwarfs consistent with the
 observations (see Fig.\ \ref{figfeh}). The properties of the stars and the
 stellar subpopulations are determined by population synthesis calculations (see
 Sect.\ \ref{pop}). 

The properties of main  sequence stars with lifetime $\tau$ are determined by an
appropriate weighted average over the age range. 
The thickness $h_\mathrm{ms}$,
the normalized density profile $\rho_\mathrm{ms}(z)/\rho_\mathrm{ms,0}$ 
and the velocity dispersion $\sigma_\mathrm{ms}$ are given by
\bqn
h_\mathrm{ms}^{-1}&=&\frac{1}{S_\tau}
	\int_0^{\tau}\frac{SFR(\tau)}{h_\mathrm{p}(t)}\dd t
	\\&&\qquad \mathrm{with}\quad S_\tau=\int_0^{\tau}SFR(\tau)\dd t\\
\frac{\rho_\mathrm{ms}(z)}{\rho_\mathrm{ms,0}}&=&
\frac{1}{S_\tau}\int_0^{\tau}h_\mathrm{ms}
	\frac{SFR(\tau)}{h_\mathrm{p}(t)}
	\exp\left( \frac{-\Phi(z)}{\sigma^2_\mathrm{W}(t)}\right)\dd t \\
\sigma^2_\mathrm{ms}(z)&=&
\int_0^{\tau}\sigma^2_\mathrm{W}(t)\frac{SFR(\tau)}{h_\mathrm{p}(t)}
	\exp\left( \frac{-\Phi(z)}{\sigma^2_\mathrm{W}(t)}\right)\dd t
	\times\\&&\left[
\int_0^{\tau}\frac{SFR(\tau)}{h_\mathrm{p}(t)}
	\exp\left( \frac{-\Phi(z)}{\sigma^2_\mathrm{W}(t)}\right)\dd t 
	\right]^{-1}\quad.
\eqn
Density profiles and thicknesses of the final model are shown in Figs.\
\ref{figrho} and\ \ref{figage} and the velocity dispersion of the main sequence stars are
compared in the lower panel of Fig.\ \ref{figsig} with the observations.

In the fitting procedure (Sect. \ref{sfr}) a pair of star formation
history and heating function is selected to derive the intrinsic structure of
the disc.
The star formation history and heating function of the final model are
shown in Fig. \ref{figsfr}.

\subsection{Gas component \label{gas}}

For the gas component we use a simple model to account for the
gravitational potential of the gas.
The vertical profile of the gas component, which is used for the gravitational 
force of the gas, is constructed dynamically 
like the stellar component. The gas distribution is modeled by
distributing the gas with a constant rate over the  velocity
dispersion range $\sigma_\mathrm{W}(t)$ of the young stars up to a maximum age
$t_\mathrm{g}$. By varying $t_\mathrm{g}$ we force the scale height of the gas 
$z_\mathrm{g}$
 to the observed value of $z_\mathrm{g}\approx 100$\,pc. 
 The surface density of the gas
 is related to the stellar surface density by the ratio 
 $Q_\mathrm{g}/Q_\mathrm{s}=\Sigma_\mathrm{g}/\Sigma_\mathrm{s}$, which is 
 chosen (together with the scale height $h_\mathrm{eff}$ of the stars)
 to be consistent with the observed surface density of 
 $\Sigma_\mathrm{g}\approx 10-13\,\msun pc^{-2}$.

\subsection{Dark matter halo \label{halo}}

The halo does not fulfil the thin disc approximation. For a spherical halo we
get the vertical component of the force to lowest order from
\bq
\frac{\dd\Phi_\mathrm{h}}{\dd z}= \frac{GM_\mathrm{R}}{R^2}\frac{z}{R}\quad, 
\eq
with $r^2=R^2+z^2$ and the enclosed halo mass $M_\mathrm{R}$ inside radius $R$. 
Comparing this with the one-dimensional Poisson equation from the thin disc
approximation (integrated over $z$ 
near the midplane to lowest order for small $z$)
\bq
\frac{\dd\Phi}{\dd z}\approx 4\pi G \rho_0 z
\eq
we can use for the local halo density 
\bq
\rho_\mathrm{h0}=\frac{M_\mathrm{R}}{4\pi R^3}
\eq
to be consistent with the spherical distribution. This value corresponds exactly 
to the singular isothermal sphere. Therefore we can
 use the thin disc approximation also for the halo by using the local halo 
 density $\rho_\mathrm{h0}$ and the halo velocity dispersion 
 $\sigma_\mathrm{h}$ estimated from the rotation curve
 by adopting an isothermal spherical halo. The effect of
 a cored halo, anisotropy and flattening, which would lead to small
 inconsistencies at large $z$, is neglected here. 
 For other halo profiles correction factors would
be necessary introducing some inconsistency in the halo profile description and
leading to a different local halo density. The latter is be more important, 
because the effect of the halo potential on the disc is stronger than the
adiabatic contraction of the halo in the disc potential. 

We use an isothermal halo component with $\sigma_\mathrm{h}\approx 100$\,km/s.
The halo surface density $\Sigma_\mathrm{h}$ is determined in the fitting
procedure, because it influences not only the total surface density
significantly, but also the shapes of the velocity
distribution functions $f(|W|)$ of the stars.


\subsection{Stellar population synthesis \label{pop}}

For the determination of luminosities, main sequence lifetimes and mass loss we
are not interpolating directly evolutionary tracks of a set of stellar masses
and metallicities. In order to get a complete coverage of stellar masses we use
instead the stellar population synthesis code PEGASE 
(Fioc \& Rocca-Volmerange \cite{pegase}) to calculate mass loss and luminosities
of ``pseudo'' simple stellar 
populations (SSPs). This means that the PEGASE code 
is used to calculate the
integrated luminosities and colour indices for a stellar population 
created in a single star-burst at 
different time-steps. These SSPs are then used to assemble
a stellar population with a given star formation history, in the sense that
the star formation history is assembled by a series of star-bursts. 
In this way we can assemble stellar populations for varying $SFR$ and metal
enrichment $\mathrm{[Fe/H]}$. 
Our SSPs are modeled by a constant star formation rate with a
duration of 25 Myr, the time resolution of the disc model. 
This is done
for a set of different metallicities and intermediate values from the chemical 
enrichment are modeled by linear interpolation.

\begin{figure}[t]
\centerline{
  \resizebox{0.98\hsize}{!}{\includegraphics[angle=270]{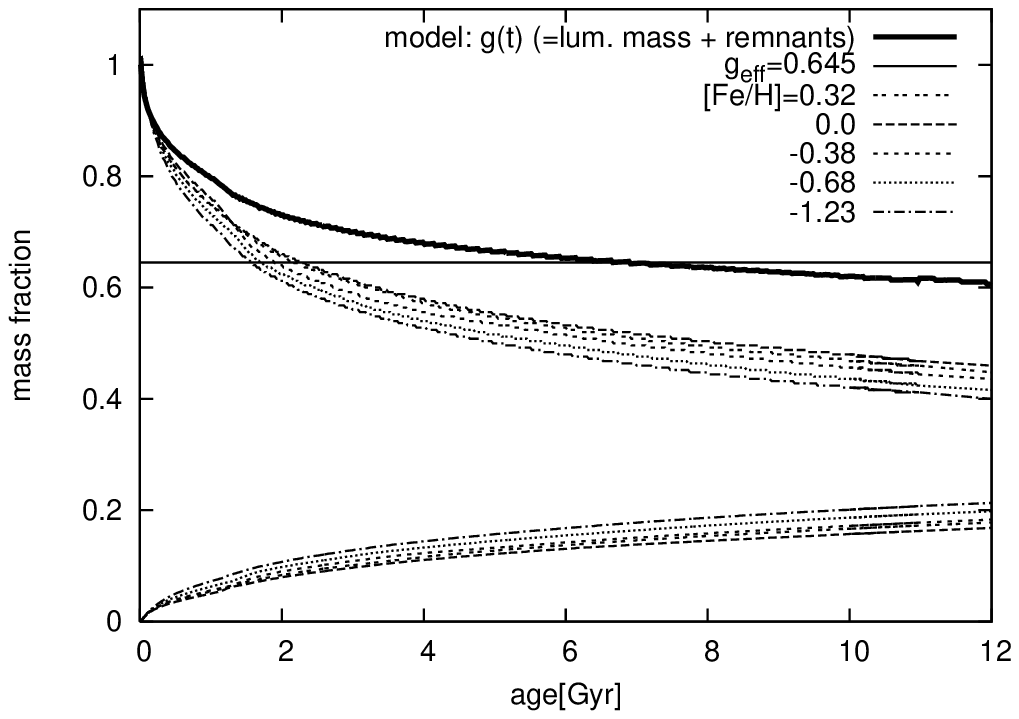}}
}
\caption[]{
Mass loss due to stellar evolution. The thin lines show the fraction of luminous
mass and of remnants (upper and lower  curve, respectively)
 for the set of metallicities used in the PEGASE code. The
thick lines shows the total stellar mass fraction (luminous plus remnants) of
the model taking into account the age-metallicity relation.
  }
\label{figmassloss}
\end{figure}

The application of the PEGASE code is twofold. On one hand mass
loss due to stellar winds, planetary nebulae and supernovae determines 
$g(t)$, the mass fraction remaining in the stellar component as a
function of age $t$. This depends on the $IMF$ and the metal
enrichment.
We adopt a Scalo-like IMF (Scalo \cite{sca86}) given by
\bqn
\dd N &\propto& M^{-\alpha}\dd M \label{eqscalo}\\
&& \alpha=\left\{\begin{array}{lcc}
        1.25&&0.08\le M/\msun < 1\\
        2.35&for&1\le M/\msun < 2\\
        3.0&&2\le M/\msun < 100
        \end{array}\right.\quad .
\eqn
and five different metallicities 
([Fe/H]= -1.23, -0.68, -0.38, 0.0, 0.32), 
which are
the input parameters into the code. Fig.\ \ref{figmassloss} shows the
mass loss for the different metallicities (thin full lines). The thin lines show 
the contribution from luminous matter and stellar remnants for the set of input 
metallicities. The sum of both contributions for each metallicity vary by a few
percent only and are therefore not shown. The thick full line
shows $g(t)$, the fraction of stellar mass in the present day stellar
disc as a function of age including the chemical evolution. 
For the oldest age-bins about 40\% of
the stellar mass is lost by stellar evolution. The total
fraction of stellar mass $g_\mathrm{eff}=0.644$ is indicated by the horizontal
line.

\begin{figure}[t]
\centerline{
  \resizebox{0.98\hsize}{!}{\includegraphics[angle=0]{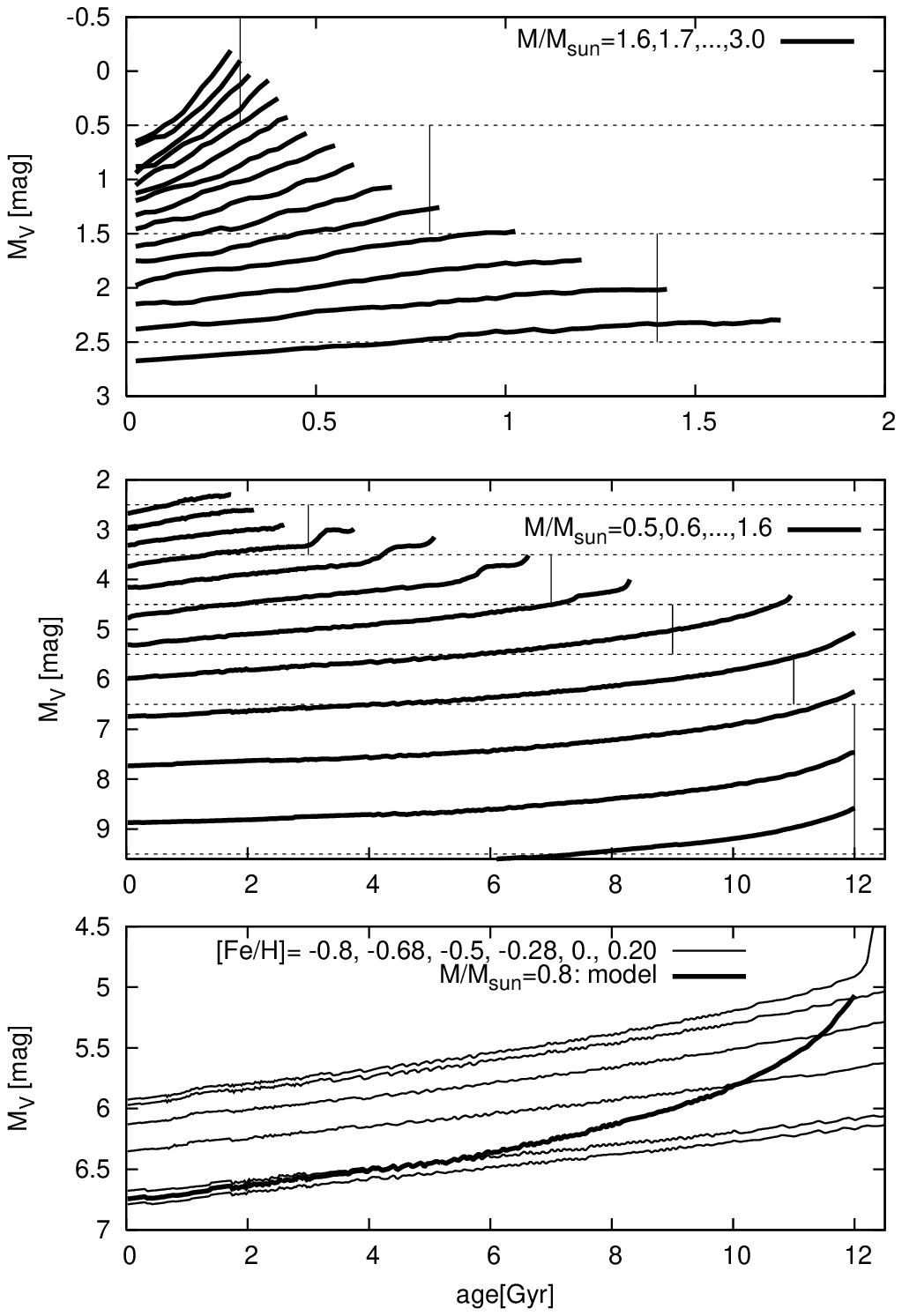}}
}
\caption[]{The lower panel shows the age dependence of the absolute luminosity 
$M_\mathrm{V}$ for stars with $M=0.8\,\msun$ (averaged over $0.1\,\msun$). Thin
lines are the evolution for different metallicities 
$\mathrm{[Fe/H]}=-0.8; -0.68; -0.5; -0.28; 0.0; 0.20; 0.32$ and the thick lines shows 
the
present day luminosity of the stars as a function of age taking into account
the age-metallicity relation.}
\label{figvage}
\end{figure}

In the second application we determine the main sequence lifetimes and
luminosities for stellar mass bins.
Here we use the PEGASE code to compute V-band luminosities of isochrones in
small mass bins.  These are needed to estimate the mean
main sequence lifetime as a 
function of the V-band luminosity $M_\mathrm{V}$ for the calculation of the
velocity distribution functions $f(|W|)$ of these stars.
We apply the PEGASE code to piecewise constant $IMF$s for mass-bins with
$0.1\msun$ and for a finer grid of metallicities 
([Fe/H]= -0.8, -0.68, -0.5, -0.28, 0.0, 0.20, 0.32). 
In Fig.\ \ref{figvage} we show $M_\mathrm{V}$ for the different mass bins as a
function of age. The lower panel gives the luminosity evolution along the main
sequence for the different metallicities (thin lines) and the age dependence of
the stars in the disc model for $M=0.8\,\msun$ demonstrating the significant
variation over the whole age range. The upper panels show the age dependent
luminosities for all mass bins. These are again not the luminosity
evolutions of the stars but the present day properties of the stellar population
taking into account the age-metallicity relation. The vertical lines indicate the estimated mean
main sequence lifetimes in the corresponding luminosity bins. The criteria to
choose these lifetimes for the determination of 
the velocity distribution functions $f(|W|)$ in the magnitude bins are discussed
in Sect.\ \ref{kinematics}.

\section{Observational Data \label{data}}

Each stellar subpopulation, which is on average older than $10^8$\,yr, is
in dynamical equilibrium with respect to the vertical distribution. Therefore
they can be used as independent samples to determine the total vertical 
gravitational potential $\phi(z)$. 
The velocity distribution functions $f(|W|)$ of the samples depend on the
age distribution of the stars. Therefore we use main sequence stars, where
$f(|W|)$ is a function of the lifetime of the stars. In order to determine the
vertical velocity distribution function $f(|W|)$ in the solar neighbourhood we
need kinematically unbiased samples with space velocities. 

\begin{figure} 
\epsfig{file=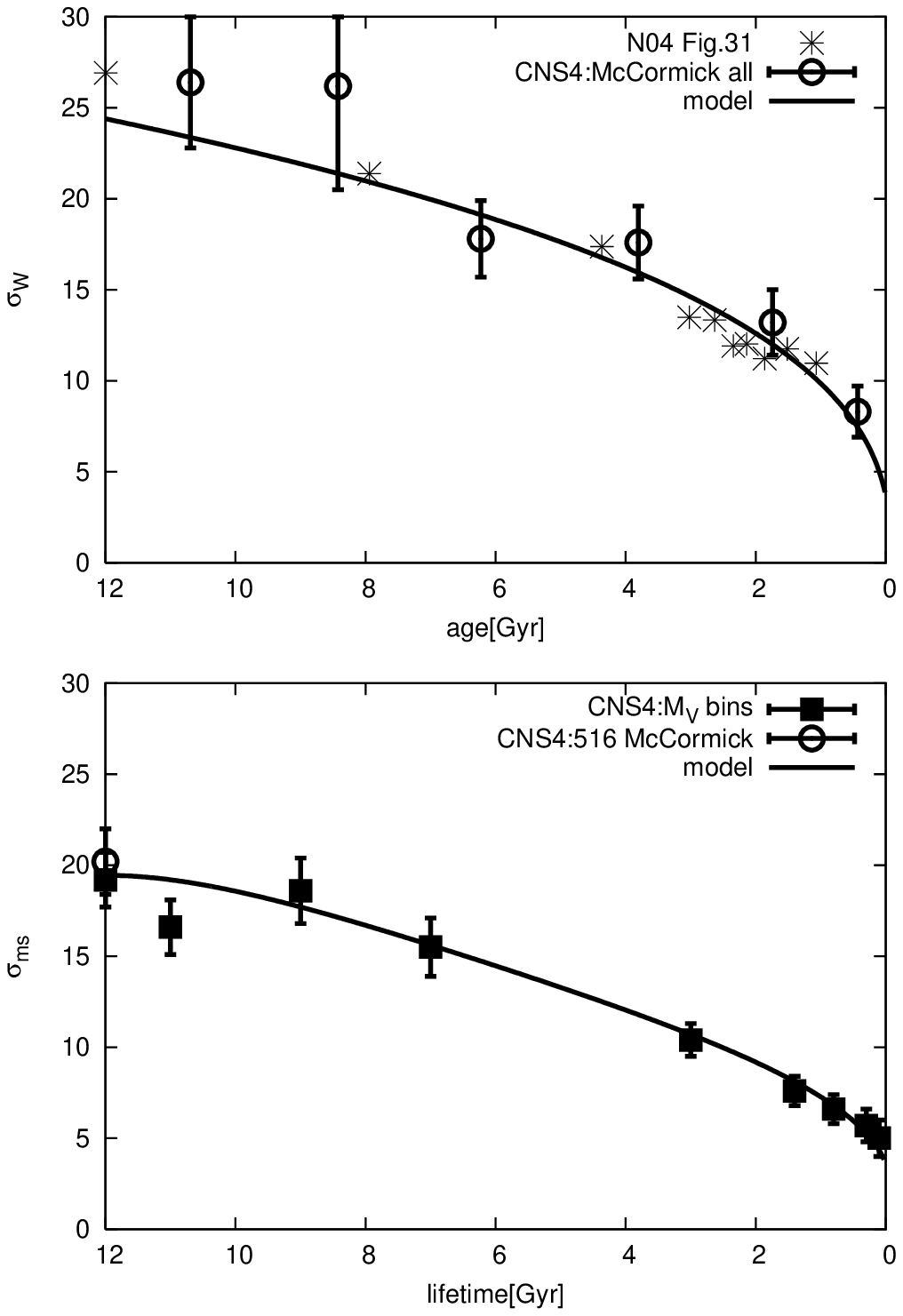,width=8.8cm,angle=0}
\caption[]{Upper panel: Velocity dispersion as a function of age for the stellar
subpopulations. Full symbols are the age groups of the McCormick K and M dwarfs. The
asterisks reproduce the age bins of Fig.\ 31 in Nordstr{\"o}m et al. (\cite{nor04}).
The full line shows the heating function AVR of the final model.
Lower panel: The symbols show the velocity dispersion of the stars along the
main sequence (magnitude bins) as a function of mean lifetime. The full line
gives the result of the final model.}
\label{figsig}
\end{figure}

\begin{figure*} 
\epsfig{file=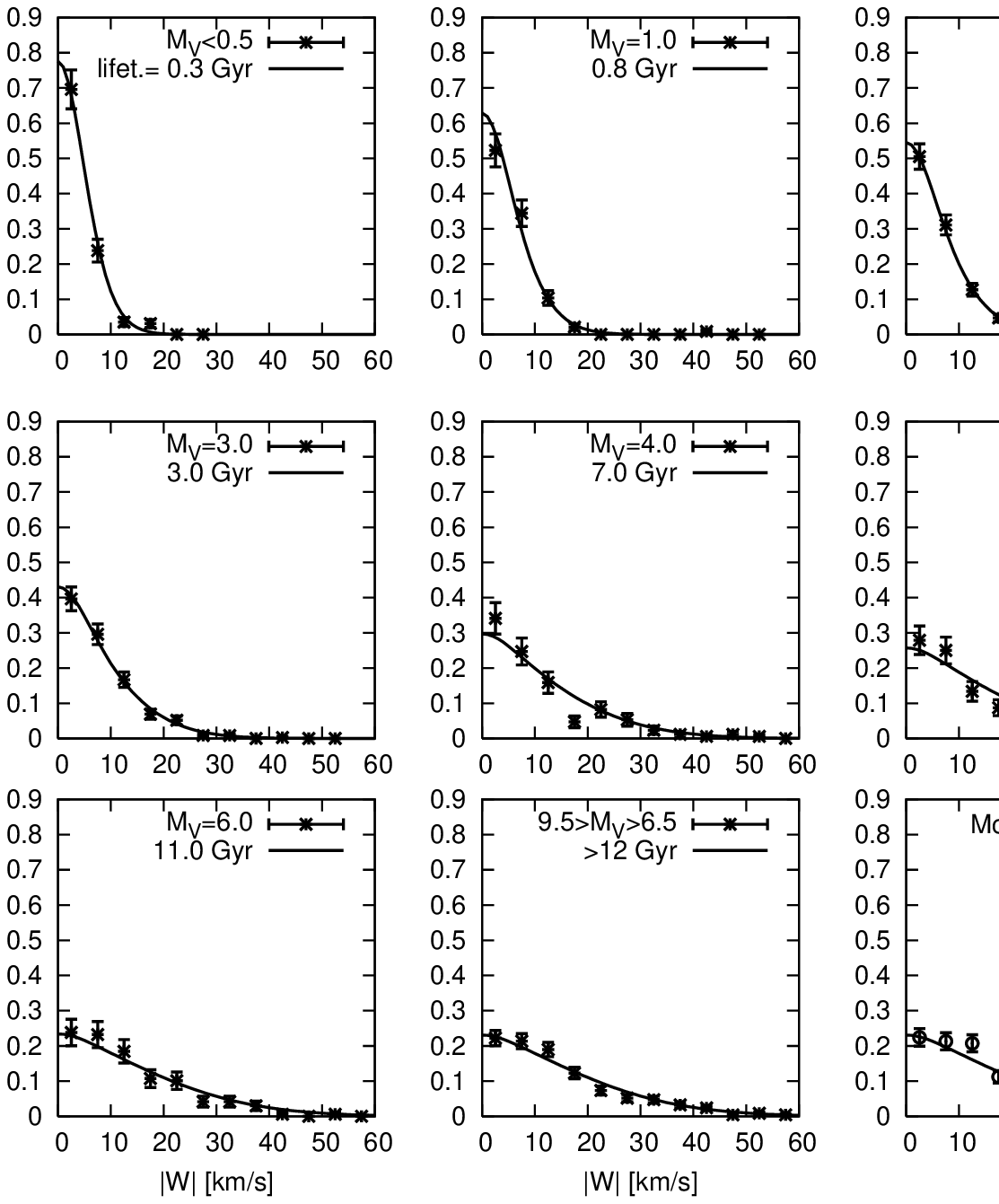,width=13cm,angle=0}
\caption[]{The normalized velocity distribution functions $f(|W|)$ for the 
different sub-samples given in Table \ref{tabdata} and for the McCormick stars 
(lower right). Symbols are the data and lines represent the model.}
\label{figfw}
\end{figure*}

For the determination of the age velocity dispersion relation we use the
McCormick K and M dwarfs (Vyssotsky \cite{vys63}). Detected by a spectroscopic survey  
they are free from kinematic bias. Altogether 516 stars show reliable distances - almost
all from the Hipparcos Catalogue - and space velocity
components. For a subsample of about 300 stars Wilson and Woolley (\cite{wil70}) estimated the 
CaII emission intensity at the H
and K lines in a relative scale allowing to construct six different age bins under the assumption of
a constant star formation rate ( see Jahrei{\ss} and Wielen, 1997).    
For each bin the vertical velocity dispersion is determined. The mean ages of the
bins are rescaled from 10\,Gyr to 12\,Gyr for the total disc age 
(see Fig.\ \ref{figsig}). 
All these stars are also used to determine the velocity distribution
function $f(|W|)$ of stars with lifetime larger than 12\,Gyr 
(see  Fig.\ \ref{figfw}). The number of stars in each individual age bin
is too low to obtain reliable $f(|W|)$.

\begin{figure} 
\epsfig{file=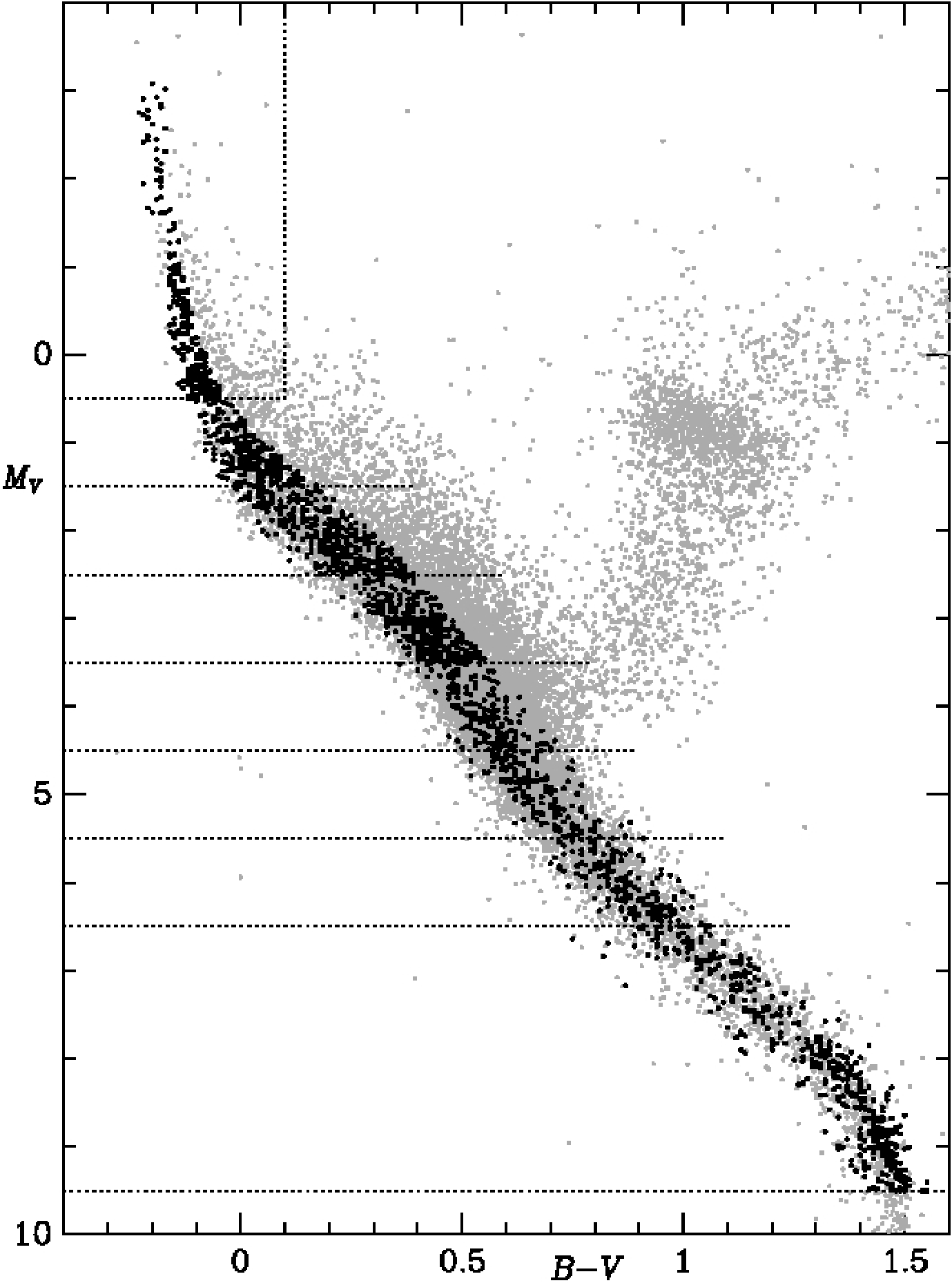,width=8.8cm,angle=0}
\caption[]{The HR-diagram shows all Hipparcos stars (grey dots) 
with $\sigma_{\pi}/\pi \le .10 $
The selected main sequence 
stars are overplotted by larger black dots and the eight
magnitude bins are indicated by the dotted lines.
}
\label{fighr}
\end{figure}

For the determination of $f(|W|)$ along the main sequence we use the Hipparcos
stars with good space velocities supplemented at the faint end down to 
$M_\mathrm{V}=9.5$\,mag by stars from an updated version of the 
Catalogue of Nearby stars (CNS4, Jahrei{\ss} et al. \cite{jah97}), i.e. also most of 
the CNS4 data rely on Hipparcos results. 
 For the determination of the space velocities good distances, proper
motions and radial velocities are required. This was achieved in combining the Hipparcos
data with radial velocities originating from an unpublished compilation of the "best" radial 
velocities for the nearby stars, which was then extended to all Hipparcos stars.

The selection process is a following. The visual absolute magnitude of the Hipparcos stars was
determined from the visual magnitude and the trigonometric parallax given in the Hipparcos
catalog. In the case of binaries resolved by Hipparcos the $ M_V $ of the brighter component was chosen
calculated from the combined magnitude and the magnitude difference measured by Hipparcos.
Then we select in the HR-diagram 
(Fig.\ \ref{fighr}) a rough regime along the main sequence in order to exclude
most of the giants and white dwarfs. After determination of the best fit for the mean
main sequence $M_\mathrm{V,ms}(B-V)$ (thick full line) all stars  in the
magnitude range $M_\mathrm{V,ms}\pm 0.8$\,mag were selected.
 Only for the brightest magnitude bin with
$M_\mathrm{V}<0.5$\,mag we use a colour cut $B-V \le 0.1$\,mag instead. Now the main
sequence is divided into magnitude bins 
$M_\mathrm{V}<0.5, 1\pm0.5, ..., 6\pm0.5, 8\pm1.5$\,mag.

In order to avoid a kinematic bias we restrict the distances of the stars in
each magnitude bin to be well within the completeness limit of the Hipparcos catalogue
 determined by the magnitude limit $ V \sim 7.3$\,mag  of the Hipparcos Survey. 
The properties of the resulting samples of main sequence stars are
collected in Table\ \ref{tabdata}. Column 1 lists the source catalogue. Column 2, 3 and 4 list  the range
 in absolute visual magnitude, the selected distance limit, and the total number of stars available, respectively.
 Column 5 and 6 list the number of
stars removed due to poor parallaxes ($\sigma_{\pi}/\pi > .15 $) and unknown or poor radial 
velocities. Finally, in column 7 the
remaining number of main sequence stars with good space velocity components is given. 

\begin{table}[h]
\caption {Complete samples of nearby main sequence stars}
\begin{center}
\begin{tabular*}{8.8cm}{l@{\extracolsep\fill}rrrrrr}
\hline 
source & $M_{V}$  & $d_{lim}$ &  N &   n$^*$ & no RV   &  N$_{fin}$\\
      &    & [pc]      &    &         &         &           \\
\hline 
Hip &$\le$-0.5& 200 &    93 &    7 &   0   & 86 \\
Hip &  0 & 200      &   183 &   27 &  15   & 141 \\
Hip &  1 & 100      &   242 &    0 &   1   & 241 \\
Hip &  2 &  75      &   401 &    8 &   7   & 386 \\
Hip &  3 &  50      &   352 &    0 &   4   & 348 \\
Hip &  4 &  30      &   172 &    2 &   0   & 170 \\
Hip &  5 &  25      &   172 &    0 &   0   & 172 \\
Hip &  6 &  25      &   170 &    0 &   2   & 168 \\
CNS & $\ge$6.5 &  25  &   525 &      &  84   & 441 \\
\hline
\end{tabular*}
\end{center}
(*) n = number of stars with $\sigma_{\pi}/\pi > .15 $
\label{tabdata}
\end{table}

For the determination of the velocity distribution functions $f(|W|)$ we correct
for the peculiar motion of the Sun using $W_\odot=7$\,km/s. The resulting
normalized distribution functions are shown in Fig.\ \ref{figfw} 
with a binning of 5\,km/s in $|W|$.

\section{Properties of the disc \label{disc}}

The disc model has some free parameters additionally to the main
input functions $SFR$ and AVR. 
In all models we fix the local stellar density to 
$\rho_\mathrm{s0}=0.039\msun pc^{-3}$ from the CNS4
and the velocity dispersion of the Dark
matter halo to $\sigma_\mathrm{h}=4\sigma_\mathrm{e}$.
For each pair of $SFR$ and AVR we determine iteratively the exponential scale
height of the stellar disc $z_\mathrm{s}$, 
the relative fractions $Q_\mathrm{s},Q_\mathrm{g},Q_\mathrm{h}$ of the surface
density (at $|z|<z_\mathrm{max}$, where $z_\mathrm{max}$ is indirectly 
determined by the depth of the potential well),
and the scale height of the gas component by choosing a 
maximum 'age' $t_\mathrm{g}$. The $Q$-values together with $z_\mathrm{s}$ determine the
maximum velocity dispersion $\sigma_\mathrm{e}$ of the stars and the surface
densities of the components. 
A comparison with the observational constraints for the surface densities and
velocity dispersions restrict the exponetial scale hwight to 
$z_\mathrm{s}$=270\,pc with an uncertainty of 10\%. 
In the final iteration the metal enrichment $\mathrm{[Fe/H]}(t)$ is adapted
to be consistent with the local metallicity distribution of G dwarfs. 
A list of model parameters and derived physical
quantities is given in Table \ref{tabmod} together with the corresponding data 
of other authors.

\begin{table}[h]
\caption {Disc properties: mean and present day $SFR$ (in $\msun/pc^2Gyr$)
are for the thin disc; 
local densities for stellar thin disc, gas, DM halo,
thick disc (indices s, g, h, t; in $\msun/pc^3$);
surface densities are below $|z_\mathrm{max}|=2.3$\,kpc
or up to 0.35, 1.1\,kpc (in $\msun/pc^2$);
(half-)thickness of the thin disc $h_\mathrm{eff}$ and exponential scale heights
$z$ are in pc;
velocity dispersions in km/s;
The second column represents the thin disc model and the third column the
values with the thick disc 1 included. 
Columns 4 and 5 give some parameters from the
literature. }
\begin{center}
\begin{tabular*}{8.8cm}{l|c|r|rr}
\hline 
quantity & thin   & + thick &other & sources\\
      &   disc       & disc &     \\
\hline 
$\langle SFR \rangle$	 & 4.2   & 3.9   &   \\
$SFR_\mathrm{p}$ 	 & 0.84  & 0.79   &    \\
$\rho_\mathrm{0}$ 	 & 0.094 & 0.092 &0.076$^1$ & 0.098$^2$ 	\\
$\rho_\mathrm{s0}$ 	 & 0.039 & 0.036 &0.045$^1$ & 0.044$^2$	\\
$\rho_\mathrm{g0}$ 	 & 0.042 & 0.041 &0.021$^1$ & 0.050$^2$    \\
$\rho_\mathrm{h0}$ 	 & 0.013 & 0.012 &0.01$^1$   \\
$\rho_\mathrm{t0}$ 	 &       & 0.003  &   \\
$\Sigma_\mathrm{s}$ 	 & 32.6  & 30.3   & 34.4$^2$    \\
$\Sigma_\mathrm{g}$ 	 & 13.4  & 13.3 &6$^1$ & 13$^2$   \\
$\Sigma_\mathrm{t}$ 	 &       & 5.7  &   \\
$\Sigma_\mathrm{disc}$   & 46.0  & 49.3  & 56$^3$ & 48$^4$  \\
$\Sigma(<0.35)$  	 & 43.2  & 42.4   &41$^3$   \\
$\Sigma(<1.1)$ 		 & 72.0  & 71.5  &74$^3$ & 71$^4$  \\
$h_\mathrm{eff}$ 	 & 416   & 418    & 	\\
$z_\mathrm{s}$ 	 	 & 270   & 270    & 	\\
$z_\mathrm{g}$ 		 & 100   & 101 &140$^1$    \\
$z_\mathrm{t}$ 		 &       & 745   &   \\
$\sigma_\mathrm{e}$ 	 & 24.4  & 24.3 &17.5$^1$    \\
$\sigma_\mathrm{h}$ 	 & 97.5  & 97.1 &85$^1$   \\
$\sigma_\mathrm{t}$ 	 &       & 41.3	 &    \\
\hline
\end{tabular*}
\end{center}
$^1$ Robin et al. \cite{rob03}; 
$^2$ Holmberg \& Flynn \cite{hol00}; 
$^3$ Holmberg \& Flynn \cite{hol04}; 
$^4$ Kuiken \& Gilmore \cite{kui91}
\label{tabmod}
\end{table}

In the next subsections we discuss the main input 
functions $SFR$ and AVR and
other properties of the disc model including a
comparison with other work and predictions for future observations. 

\subsection{Star formation history and dynamical heating \label{sfr}}

For a given stellar density profile there is a series of
function pairs ($SFR$,AVR) to construct a dynamical equilibrium model.
Strong additional constraints originate in the correlation of the
velocity distribution functions $f(|W|)$ of
stars with different lifetimes. This leads to restrictions for the AVR and in
turn to the $SFR$. The sequence of   $f(|W|)$ along the main
sequence is a direct measure of the age distribution of stars in the local
volume, which are converted to surface densities for connecting to the
$SFR$.

In the final model we use the star formation history (see Fig.\ \ref{figsfr})
\bqn
SFR(\tau)&=&S_0\frac{a(\tau+t_0)}{(\tau^2+b^2)^2}\quad\mathrm{with}\\
a&=&34.5\, ; \quad b=4.2\, ; \quad t_0=0.01\,,
\eqn
where the time $\tau$ is in units of Gyr and $S_0=50.5\,\msun/pc^2$ is 
connected to the stellar surface density by Eq.\ \ref{eqSigmas}. The $SFR$
decreases by a factor of 10 from the maximum to the present day value, which
corresponds to an e-folding timescale of 5\,Gyr. The present day star formation
rate with 20\% of the average value of 4.1\,$\msun/pc^2 Gyr$ is relatively low.
It is confirmed by the $IMF$ considerations (see Sect.\ \ref{imf}).

For the dynamical heating function AVR (see Fig.\ \ref{figsig}) 
we use a power law
\bqn
\sigma_\mathrm{W}(t)&=&\sigma_\mathrm{e}
 \left(\frac{t+t_0}{t_\mathrm{a}+t_0}\right)^\alpha\quad\mathrm{with}\\
\alpha&=&0.375\,  ; \quad t_0=0.0767\,,
\eqn
where $t$ is the age in Gyr. The velocity dispersion of newly born stars is
$\sigma_\mathrm{p}=0.15\sigma_\mathrm{e}$ and $\sigma_\mathrm{e}=24.4$\,km/s is
the maximum velocity dispersion of the oldest stars with an age of 
$t_\mathrm{a}=12$\,Gyr. 
The power law index $\alpha=0.375$ is in the range of the classical
value of 1/2 from Wielen (\cite{wie77}), of 0.47 (Nordstr{\"o}m et al.
\cite{nor04}) and the value of 1/3 derived by Binney et al. (\cite{bin00})
from proper motion measurements.

\subsection{Density profiles \label{density}}

The vertical density profiles of the gas and of the stars are flattened at the
galactic plane. A measure of the flattening is given by the ratio of the
thickness and the exponential scale height. 

The corresponding stellar disc scale height is $z_\mathrm{s}=270$\,pc (from 
Eq.\ \ref{eqze} with $C_\mathrm{z}=1.18$), whereas the half-thickness
$h_\mathrm{eff}$ is 50\% larger.

For the gas, the thickness is
160\,pc compared to the scale height of 100\,pc. The density profile and 
the surface density is consistent with the observed HI-profile (Dickey \& Lockman
(\cite{dic90}) taking into account about 50\% of H$_2$ 
(Bronfman et al. (\cite{bro88}) within the large uncertainty range.

Due to the gravitational potential of the disc, the local density of the dark matter halo is
55\% larger than the DM density at $z=2.3$\,kpc and 18\% larger than the 
mean DM density.
\begin{figure} 
\epsfig{file=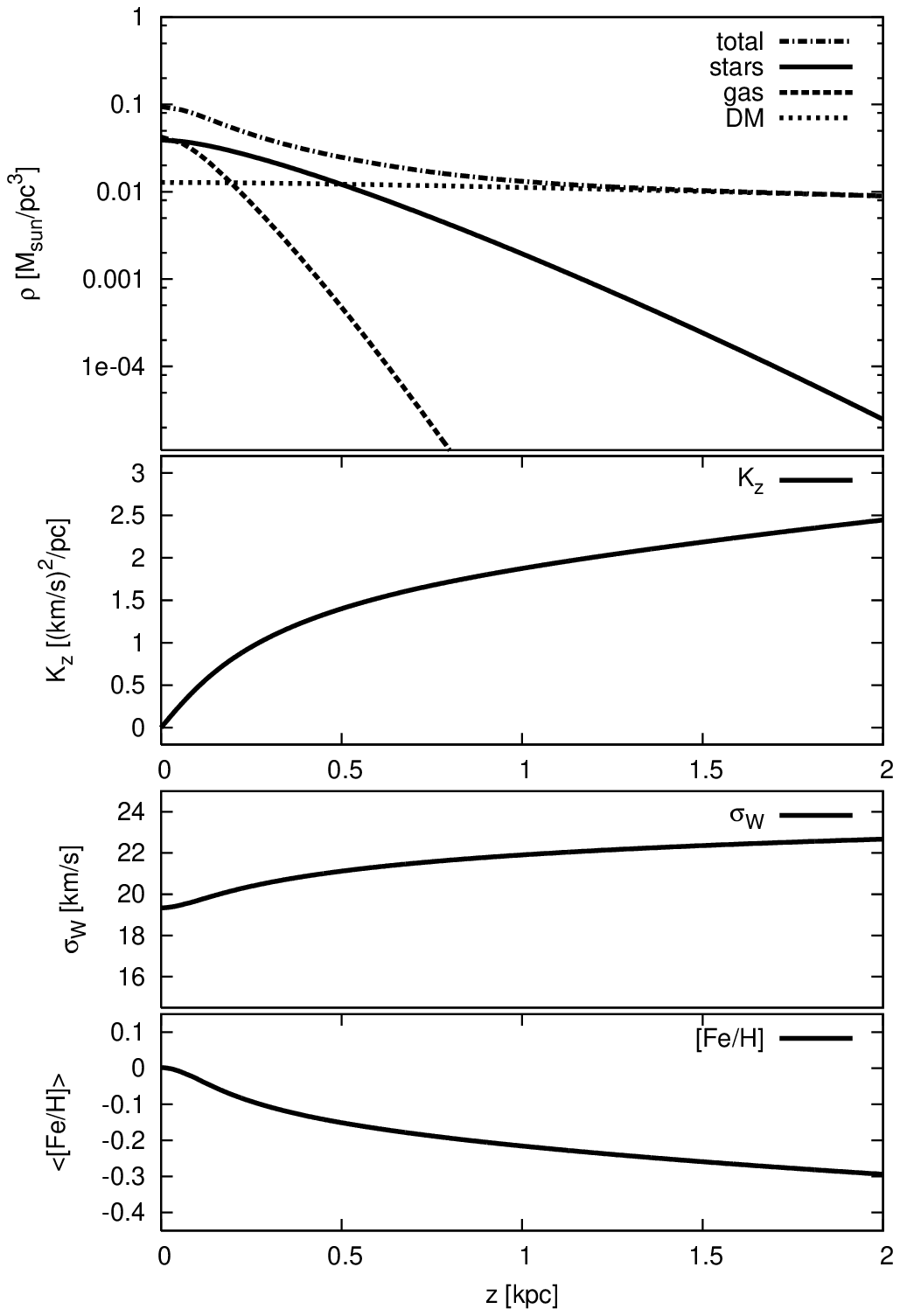,width=8.8cm,angle=0}
\caption[]{Vertical profiles of the disc. From top to bottom: 
Density profiles of the components; $K_\mathrm{z}$ force, which is proportional
to the cumulative surface density;
velocity dispersion of the stellar component;
V-band luminosity weighted mean metallicity of the stellar component.}
\label{figrhoz}
\end{figure}

\begin{figure} 
\epsfig{file=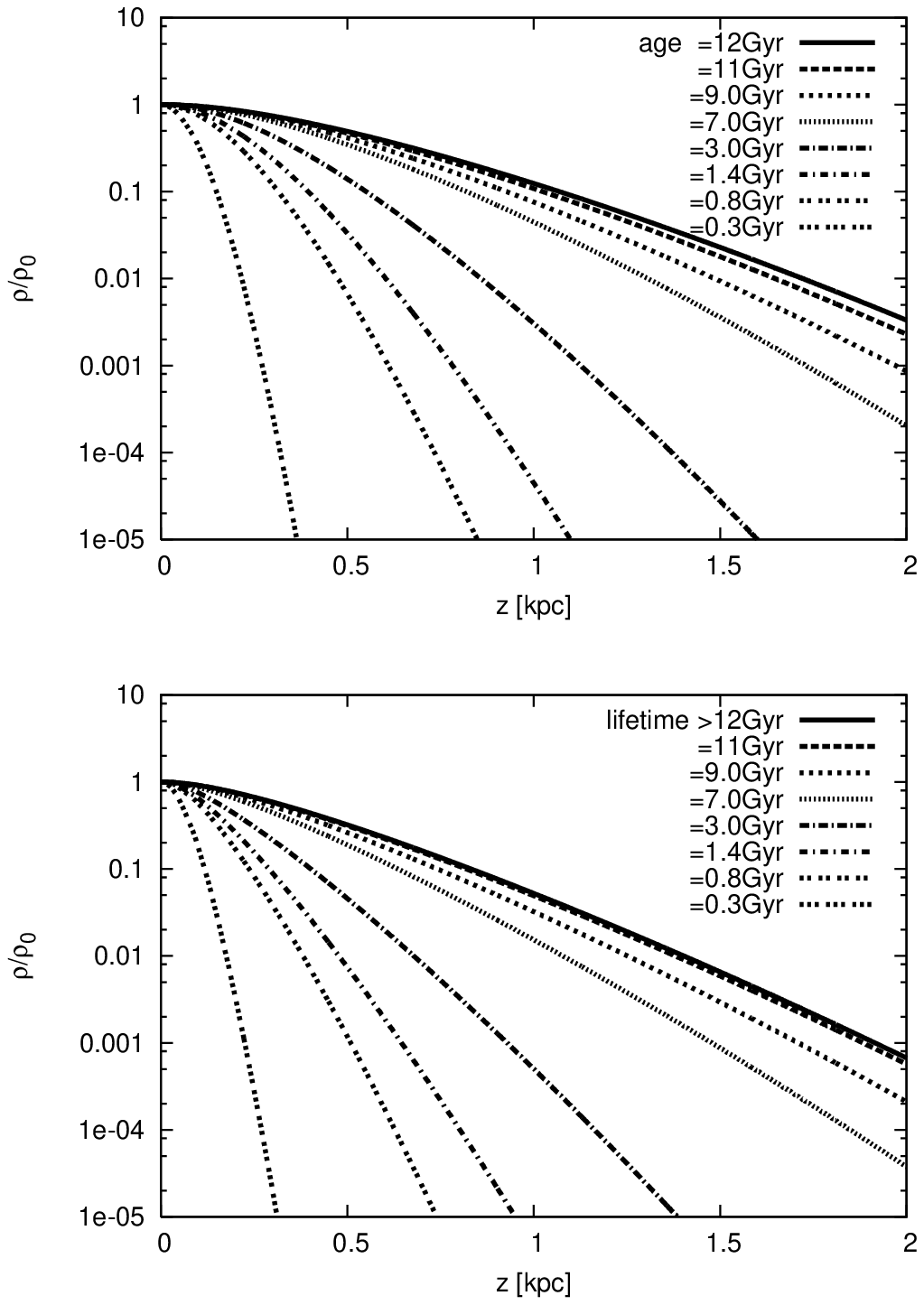,width=8.8cm,angle=0}
\caption[]{Upper panel: Normalized density profiles of the stellar
subpopulations as a function of age.
Lower panel: Same for the main sequence stars as a function of lifetime. The set
of lifetimes corresponds to the mean lifetimes of the subsamples used for the
velocity distribution functions (see fig.\ \ref{figfw}).}
\label{figrho}
\end{figure}
The density profiles of main sequence stars differ in shape from the profiles of
the subpopulations for a given age. The lower panel of Fig.\ \ref{figrho} shows
the normalized density profiles with lifetimes according to the samples used for
 the model. They are
steeper than the corresponding profiles of the subpopulations with the same age
and significantly shallower than the density profile using the mean age of the
subpopulation. This difference is quantified by the thicknesses $h_\mathrm{p}$ and
$h_\mathrm{ms}$ shown in the lower panel of Fig. \ref{figage}.

The density profiles derived by
Holmberg and Flynn (\cite{hol00}) for two sub-samples of A stars (with 
$0<M_{V}<1$\,mag) and early F stars (with $1<M_{V}<2.5$\,mag) are in full
agreement with the corresponding profiles with lifetimes of 0.3 and 0.8\,Gyr in
our model (see Fig.\ \ref{figrho}). The $K_\mathrm{z}$ force of their reference
model is also in good agreement with our results (see Fig.\ \ref{figrhoz}).

\subsection{Age distributions \label{ages}}

\begin{figure} 
\epsfig{file=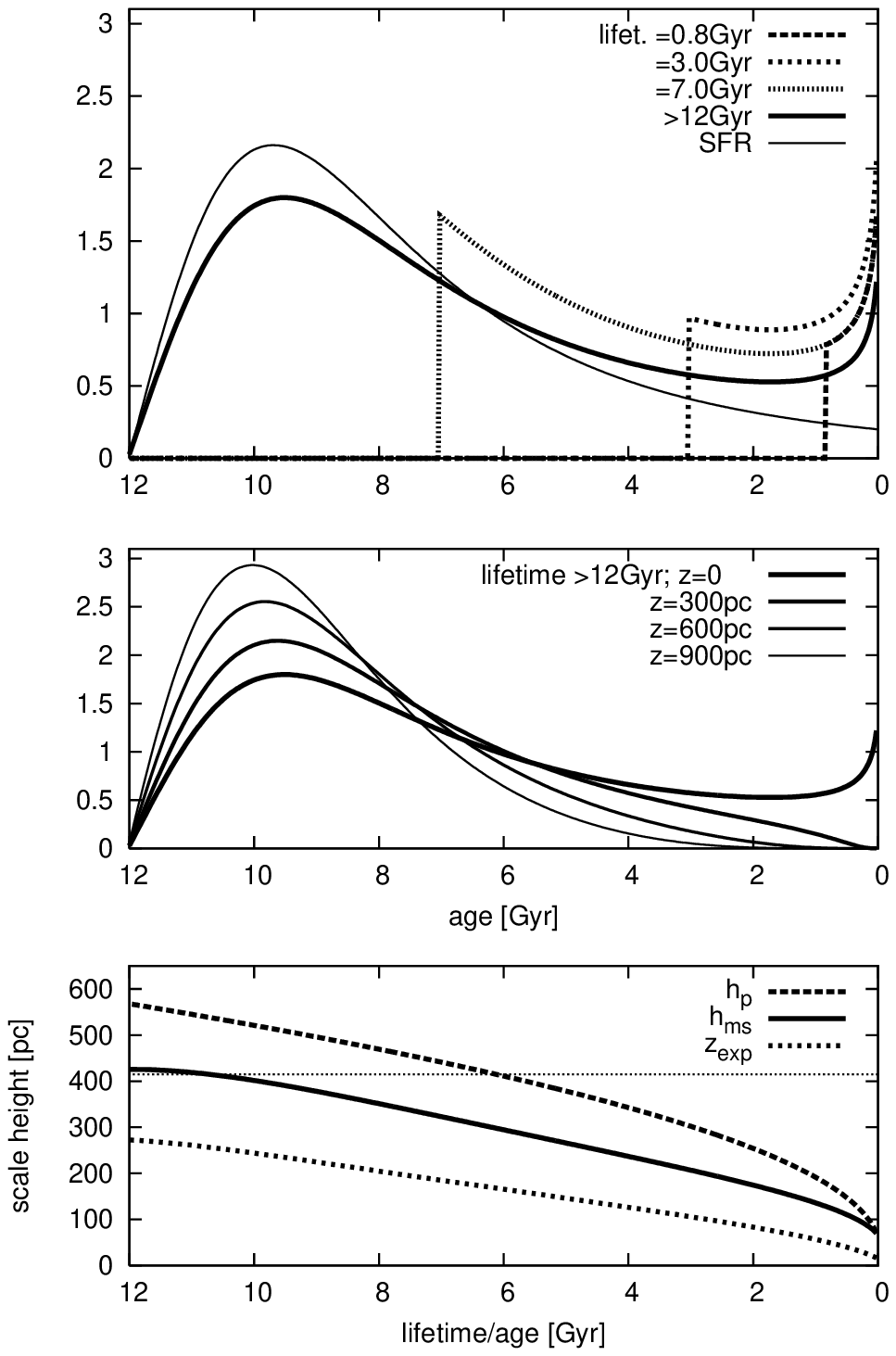,width=8.8cm,angle=0}
\caption[]{The upper panel shows the normalized local age distributions of 
main sequence stars with different lifetimes. The cumulative mass fraction of
all stars and the $SFR$ are also plotted.
The middle panel shows the age distribution of K and M dwarfs and the cumulative
stellar mass fraction with increasing height $z$ above the plane.
The lower panel shows the (half-)thickness of the subpopulations as a function
of age $h_\mathrm{p}$ and of the main sequence stars as a function of lifetime
 $h_\mathrm{ms}$. For comparison the exponential scaleheights $z_\mathrm{exp}$
 for the main sequence stars are included. The horizontal line is the overall
 thickness of all stars $h_\mathrm{eff}$ (Eq.\ \ref{eqrhos0})}
\label{figage}
\end{figure}

The age distributions of the stellar sub-samples of MS stars depend on the
lifetime and on the vertical distribution. Fig.\ \ref{figage} shows in the upper
panel the age distribution of stars along the man sequence. Models of the
star formation history, which correct for the finite lifetime only and
 do not take into account the scale height 
variation, determine the local age distribution of K and M dwarfs. It varies by
less than a factor of two around the mean value. Binney et al.\ (\cite{bin00})
and Cignoni et al.\ (\cite{cig06}) propose a constant local age distribution in
this sense. Taken the uncertainties of isochrone ages into account, this is
consistent with our model.
 
The effect of the scale height correction is demonstrated by including the
$SFR$ (thin line), which represents the age distribution of K and M
dwarfs vertically integrated over the disc. Especially for young stars there is
a significant overrepresentation in the solar  neighbourhood.
In order to quantify the correction to the surface density, the lower panel of
Fig.\ \ref{figage} shows the (half-)thickness $h_\mathrm{ms}$ as a function of lifetime compared
to the exponential scale height of the same populations $z_\mathrm{exp}$ and to
the thickness of the stellar subpopulation as a function of age $h_\mathrm{p}$.

The age distributions are a strong function of $z$ above the plane. The middle
panel of Fig.\ \ref{figage} shows the lack of young stars in steps of $\Delta
z=300$\,pc above the midplane.

\subsection{Kinematics of the disc \label{kinematics}}

The AVR can be observed only by kinematically unbiased stellar subsamples with
direct age determinations. The upper panel of Fig.\ \ref{figsig} shows
the AVR of the model compared to two data sets. The circles are the McCormick K
and M dwarfs with ages determined from H and K
line strength \cite{jah97} and the asterisks represent the F an G stars of
Nordstroem et al.\ \cite{nor04} with good age determinations.
\begin{figure} 
\epsfig{file=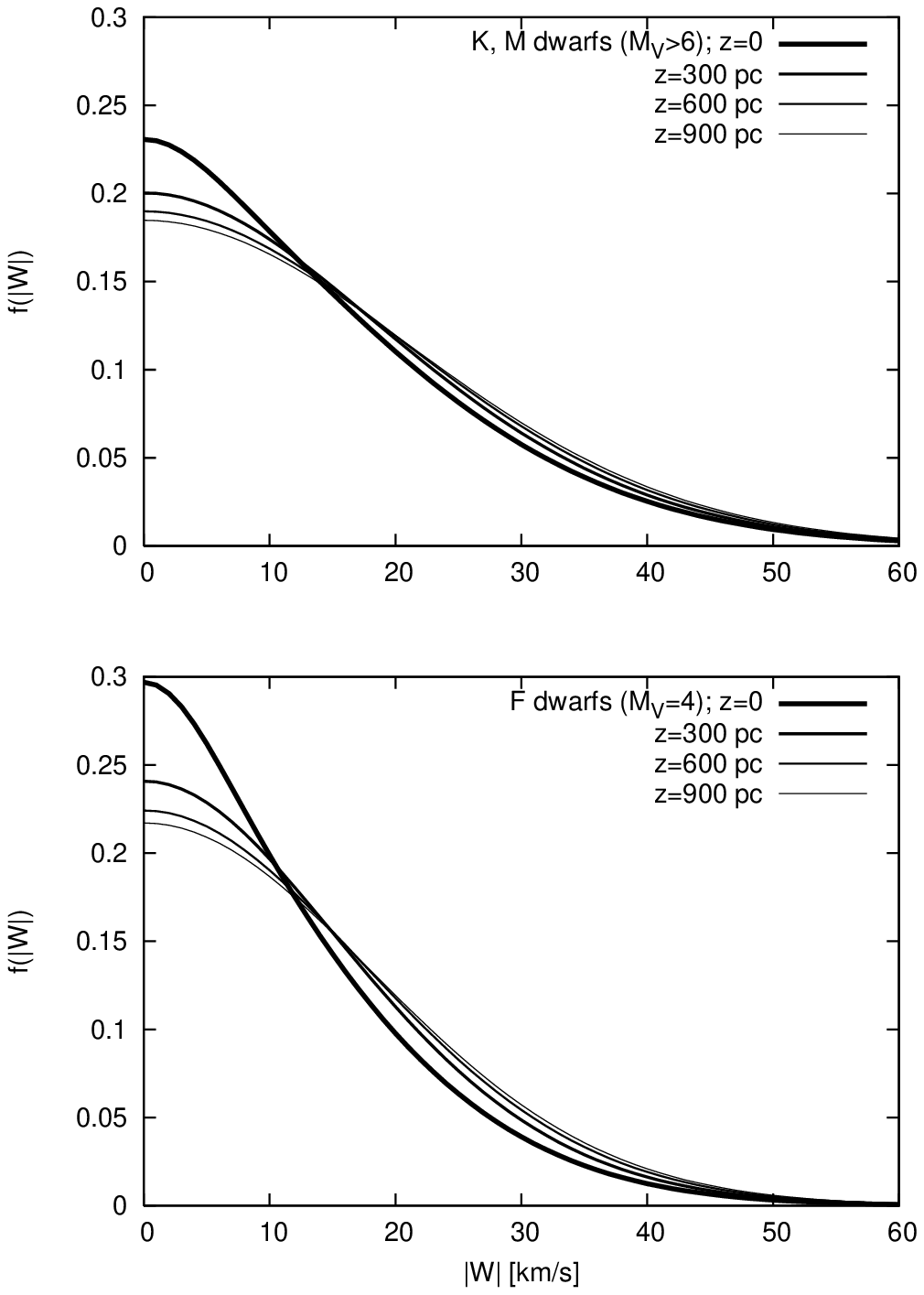,width=8.8cm,angle=0}
\caption[]{The normalized velocity distribution functions
$f(|W|)$ change with increasing $z$ (thick to thin lines) due to the varying
age mixture. 
The upper panel shows of K and M dwarfs with lifetimes larger than the age 
of the disc, whereas the lower panel is for late F dwarfs with lifetime 7\,Gyr.}
\label{figfwz}
\end{figure}

The velocity dispersions of the subsamples of main sequence stars with no age
subdivision are determined by the weighted mean over the lifetime with the local
age distribution. The lower panel of Fig.\ \ref{figsig} shows the excellent 
agreement of the model with the data from the nearby stars.
Additionally the shape of the velocity distribution functions $f(|W|)$ 
of the main sequence star samples are in good agreement with the model 
(Fig.\ \ref{figfw}). 
The shape of $f(|W|)$ depends on the lifetime and also on the
height $z$ above the midplane. Fig. \ref{figfwz} shows the
variation above the plane for K and M dwarfs (upper panel) and for F stars with
$M_\mathrm{V}=4$\,mag and a lifetime of 7\,Gyr (lower panel).

\subsection{Metallicity \label{feh}}

\begin{figure} 
\epsfig{file=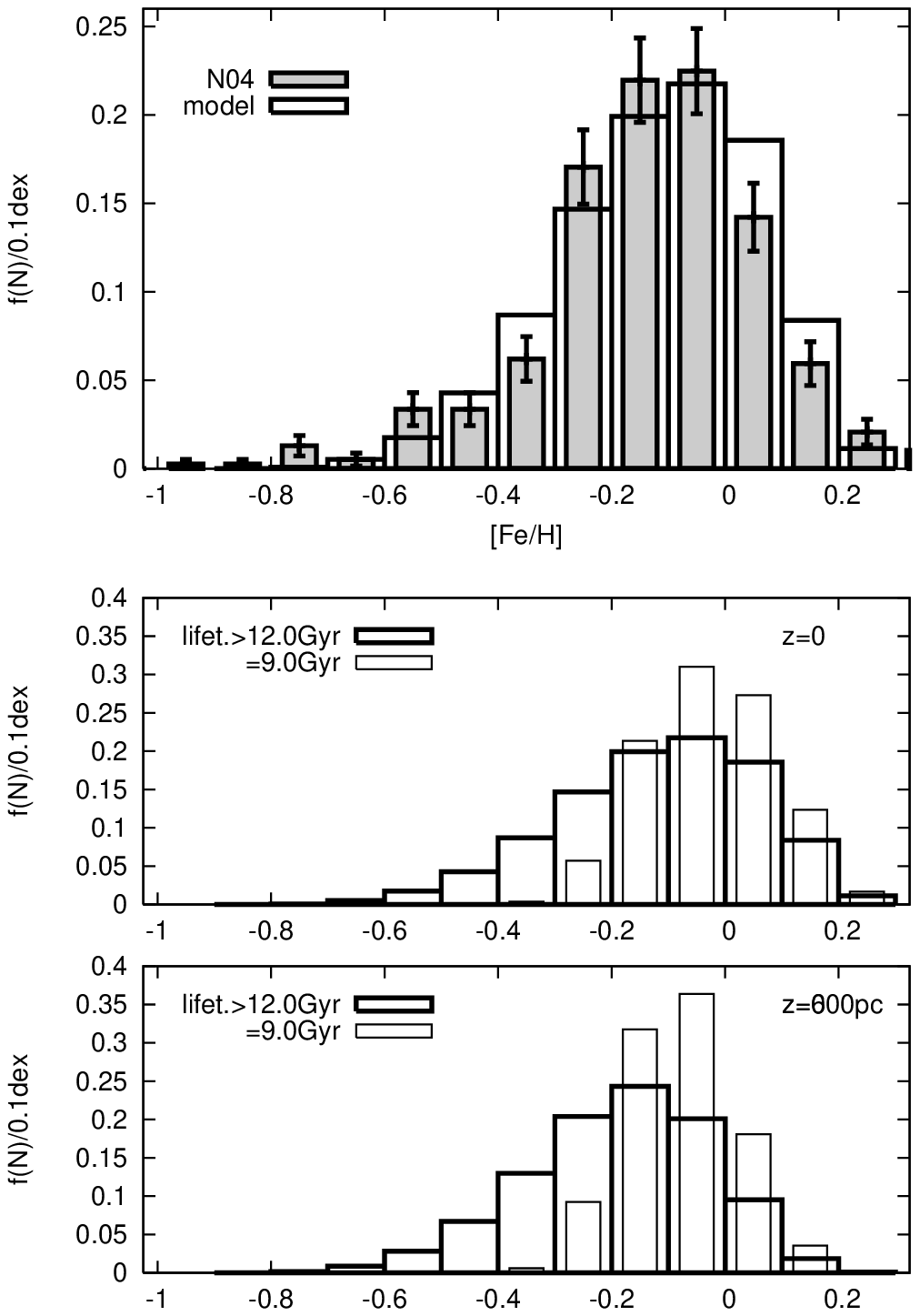,width=8.8cm,angle=0}
\caption[]{Upper panel: The metal enrichment law for the model (full line). The grey
histogram shows the observations from the Copenhagen survey for late G dwarfs
in the stellar mass range $0.84\le M/\msun\le 0.90$.
Lower panels: The metallicity distribution for early G dwarfs (lifetime 9\,Gyr;
thin lines) in
comparison to the late G, K and M dwarfs (lifetime $>12$\,Gyr; thick lines) in the
solar neighbourhood and 600\,pc above the midplane.}
\label{figfeh}
\end{figure}

We constructed an analytic metal enrichment law $\mathrm{[Fe/H]}(\tau)$, 
which reproduces the local metallicity distribution of
late G dwarfs for the
disc model with the final $SFR$ and AVR.  
The local metallicity distribution is determined from the
Copenhagen F and G  star sample (Nordstr{\"o}m et al. \cite{nor04}) selecting all
stars with masses $0.84\le M/\msun\le 0.90$ up to the completeness limit
$r<40$\,pc. The lower mass limit of $0.84\msun$ was chosen in order to avoid an overrepresentation of 
metal poor stars.
The metal enrichment law (Fig.\ \ref{figsfr}) is given by
\bqn
\mathrm{[Fe/H]}(\tau)&=&\mathrm{[Fe/H]}_0+2\mathrm{lg}\left[1+p\ln(1+q\tau)\right]
\quad\mathrm{with}\\
\mathrm{[Fe/H]}_0 &=&-0.7\, ; \quad p=0.655\, ; \quad q=0.833\,
\eqn
with time $\tau$ in Gyr. The initial metallicity
is [Fe/H]$_0$=-0.7 and the present day metallicity is 
[Fe/H]$_\mathrm{p}$=0.12. Before binning the theoretical distribution
we add an observational scatter with FWHM=0.165\,dex.
The comparison of the derived local metallicity distribution for late G dwarfs
with the data is shown in the upper panel of Fig.\ \ref{figfeh}. 
The lower panels show the predicted metallicity distribution for late
type stars (thick lines) and for late F stars with lifetime 9\,Gyr (thin lines) at
the midplane and 600\,pc above the plane.

The metal enrichment [Fe/H] and the $SFR$ can be embedded in a local chemical evolution
model. We proceed in the following way. We convert the metal abundance  
[Fe/H] to the oxygen abundance using the correlation
of [O/Fe] and [Fe/H] from Reddy et al. (\cite{red03}) by the
linear approximation [O/H]=0.375[Fe/H]. For the oxygen
enrichment we adopt instantaneous recycling and mixing to determine the infall
rate of primordial gas. The mass loss from stellar evolution is taken into
account. We start with a negligible initial amount of gas. 
Then the oxygen yield in solar units is given by 
$y_\mathrm{ox}=\langle Z_\mathrm{ox}\rangle +Z_\mathrm{ox,p}
\Sigma_\mathrm{g}/\Sigma_\mathrm{s}=1.12$ from the present day surface densities
of gas and stars, the present day oxygen abundance 
$\mathrm{lg}(Z_\mathrm{ox,p})=\mathrm{[O/H]_p}=0.045$ and the mean oxygen
abundance $\langle Z_\mathrm{ox}\rangle$ averaged over the star formation history.
The infall rate and the surface density
of gas and stars are shown in Fig.\ \ref{figsfr}. 

\subsection{Thick disc \label{thd}}

\begin{figure} 
\epsfig{file=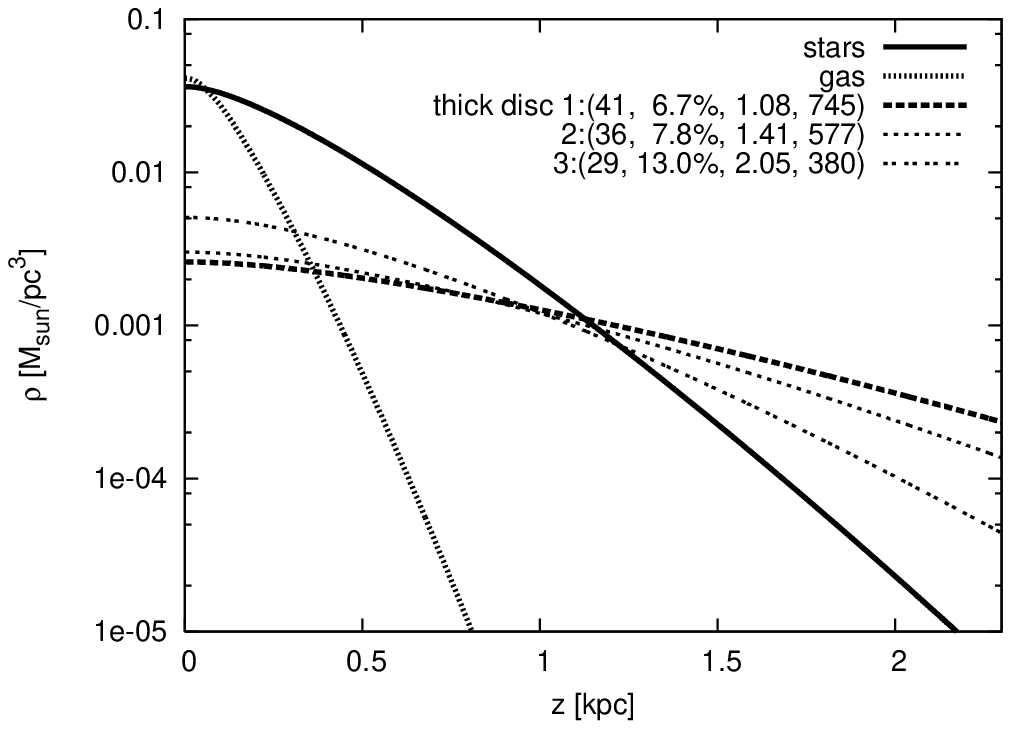,height=8.8cm,angle=270}
\caption[]{Self-consistent isothermal thick disc models with different velocity
dispersions. Stars, gas, thick disc 1 form a self-consistent model with
parameters 
($\sigma_\mathrm{t},\rho_\mathrm{0,t}/\rho_\mathrm{0,s+t},\alpha,h_\mathrm{t}$).
Velocity dispersion and local mass fraction of the thick disc are input
parameter and the power law index and exponential scale height are best fit 
parameter for a sech$^\alpha(z/\alpha h_\mathrm{t})$ profile. For comparison two
alternative cooler thick disc components are plotted.
}
\label{figthd}
\end{figure}

Since the thick disc contributes a few percent to the local density only, its
properties are not well determined in the local disc model. Therefore we derive 
here as an example the density profile of an isothermal thick disc component and discuss
the effect on the thin
disc. We split the local stellar density into a thin disc and a thick disc
component. The local mass fraction of the thick disc is determined to equal the
thin disc density at $z\approx 1.2$\,kpc as indicated by K dwarf
density profiles (Phleps et al. \cite{phl05}).
We choose the thick disc velocity dispersion and calculate the complete disc
model including the new component. The parameters of all components have to be
adjusted iteratively, because the stellar disc profile (thin plus thick disc) has a
different shape now. The resultant profiles for thin disc, gas and thick disc 
 using $\sigma_\mathrm{t}=41$\,km/s and 
$\rho_\mathrm{0,t}=2.6\times 10^{-3}\msun/\mathrm{pc}^3$ corresponding to 6.7\%
of the local stellar density are shown in Fig.\ \ref{figthd} (thick lines). The
density profile of the thick disc can be fitted by a
sech$^\alpha(z/\alpha h_\mathrm{t})$ profile to better than 3\%. 
The corrections of the
thin disc parameters due to the thick disc is of the order of 1\%. Only the
surface density of thin and thick disc is 10\% larger with a corresponding
 reduction of the Dark matter (see Table \ref{tabmod}).
 For two alternative models with smaller velocity dispersion of the thick disc
 the density profiles can also be fitted very well by the sech$^\alpha$ law. The
 parameters are also given in Fig.\ \ref{figthd}.
 
\subsection{Initial mass function \label{imf}}

\begin{figure} 
\epsfig{file=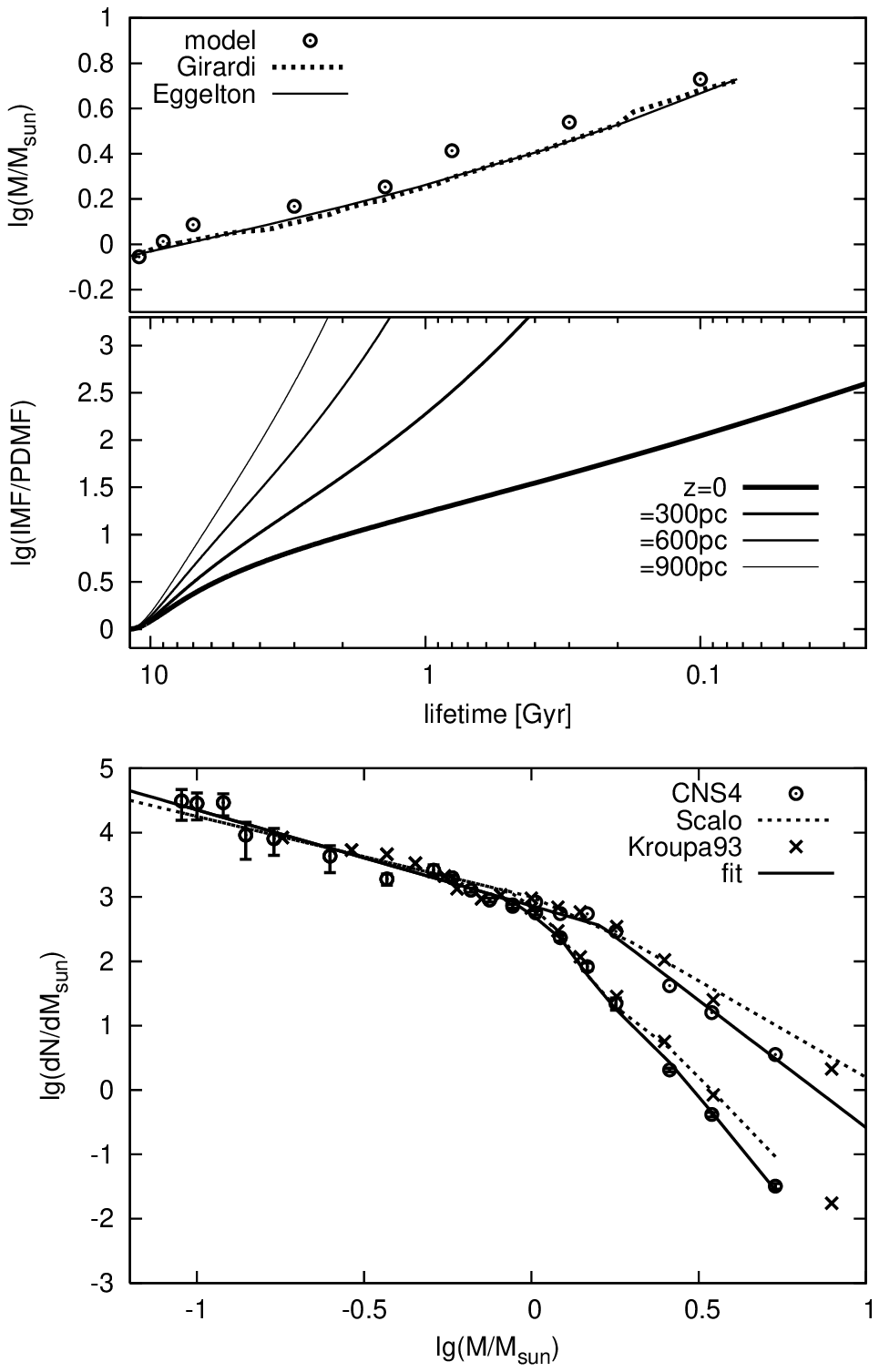,width=8.8cm,angle=0}
\caption[]{Upper panel: Stellar mass as function of lifetime for the present day
population. Symbols are the values used (from Fig. \ref{figvage}), the lines are
estimated values from evolution tracks.
Middle panel: Correction factor (=IMF/PDMF), which must be applied to 
the IMF to derive the local star counts of
stars with corresponding lifetime relative to low mass stars with lifetime
exceeding the disc age.
Lower panel: Open circles show the PDMF and the IMF of our sample from CNS4 and
Hipparcos. Crosses are the PDMF of Kroupa et al. (\cite{kro93}) and the IMF by
correcting with the same dilution factor. The dashed lines are the Scalo IMF and
PDMF used in the model and the full lines give the new best fit for the derived
IMF (open circles).}
\label{figpdmf}
\end{figure}

For the determination of the IMF from the local luminosity function, the
luminosity function is first converted to a mass function $dN=f(M)dM$ using the
transformation formulae of Henry \& Mc Carthy (\cite{hen93}), corrected in 
Henry et al. (\cite{hen99}), and extended to bright stars according to 
Schmidt-Kaler in Landolt-B\"ornstein (\cite{lan82}). 
The result is shown in the lower panel of
Fig. \ref{figpdmf} and compared to the PDMF given in Kroupa et al. 
(\cite{kro93}). The star numbers are normalized to a sphere with radius
R=20\,pc. In order to determine the correction factor converting the IMF to the
local PDMF due
to the finite lifetime and the vertical thickness $h_\mathrm{ms}$, the main
sequence lifetime is needed. We use the lifetimes estimated from the
isochrone evolution shown in Fig. \ref{figvage} for the different magnitude bins, where
the metal enrichment and the relative weighting due to the increasing number of
stars with decreasing mass in the mass interval is taken into account. These
data are shown in the upper panel of Fig. \ref{figpdmf} with a
comparison of the lifetimes determined directly from evolutionary tracks of 
Girardi et al. (\cite{gir04}) and with the analytic fitting formula of Eggleton
et al. (\cite{egg89}). Any systematic variation of the lifetimes result in
significant corrections to the correction factor and therefore the IMF.
This is the most uncertain input to the IMF determination. The correction
factors in the solar neighbourhood (z=0 thick line) and for populations above
the plane are shown in the middle panel of Fig. \ref{figpdmf}.

The used Scalo IMF (Eq. \ref{eqscalo}) is consistent with the observed PDMF. But
the kink near 1\,$\msun$ seems artificial, because it is near that mass, where
the correction factor due to the finite lifetime starts to apply.
Therefore we determined a new IMF by fitting power laws in two mass regimes
only. The best fit is
\bqn
\dd N &\propto& M^{-\alpha}\dd M \quad\mathrm{with}\\
&&\alpha=\left\{\begin{array}{rcl}
        1.49\pm 0.09&&0.08\le M/\msun < 1.61\\
	&for&\\
        3.95\pm 0.33&&1.61\le M/\msun < 6.0
        \end{array}\right.\nonumber\\
N_0&=&359/\msun\quad \mathrm{at}\quad M/\msun=1.61\,,\nonumber
\eqn
where $N_0$ gives the normalization in the 20\,pc sphere.
The back-reaction of the corrected IMF to the disc model via the mass loss is
very small and not included in the model. 

The strongest constraints on the $IMF$ at high masses and the present day $SFR$
is the observed number of A stars in the solar neighbourhood. Since the bright
stars with $M_\mathrm{V}<0.5$\,mag are observed in a sphere with a radius of
200\,pc, the sample size is a direct measure of the local surface density. Therefore
the conversion from the $IMF$ to the mean $SFR$ in the last few 100\,Myr 
depends only on the lifetime of the stars. That means, a higher present day
$SFR$ requires a shorter lifetime for the A stars or an even steeper $IMF$.

The position of the Sun is probably by about 20\,pc above the midplane
(Humphreys \& Larsen \cite{hum95}).
The local density of the subpopulations with small scale
height are significantly smaller than the midplane density
($M_\mathrm{V}<1.5$\,mag). We corrected for that implicitly, 
since we determined the
midplane density of these magnitude bins for the PDMF using spheres with large
radii (see Table \ref{tabdata}), where the offset can be neglected.

\section{Summary \label{summary}}

We presented an evolutionary disc model for the thin disc in the solar cylinder
 based on a continuous  star formation history (SFR) 
 and a continuous dynamical heating (AVR) of
the stellar subpopulations.
The vertical distribution of the stellar
subpopulations are calculated self-consistently in dynamical equilibrium. The
SFR and AVR of the stellar subpopulations are determined by
fitting the velocity distribution functions of main sequence stars.
A chemical evolution model with reasonable gas infall rate is included, 
which reproduces the local [Fe/H] distribution of G dwarfs.

We found a vertical disc model for the thin disc including the gas and dark
matter component, which is consistent with the
local kinematics of main sequence stars and fulfils the known constraints on the
surface densities and mass ratios. The SFR
shows a maximum 10\,Gyr ago declining by a factor of 10 until present time
corresponding to an e-folding timescale of 5\,Gyr.
The velocity dispersion on the upper main sequence depends on the lifetime
of the stars, which is derived from the AVR. 
For the AVR we find a power law with index of 0.375. 
Applying the stellar lifetime and the new scale height corrections to the PDMF
results in a best fit IMF with power-law indices of 
1.5 below and 4.0 above 1.6\,$\msun$, which has no kink around 1\,$\msun$.

Including a thick disc component consistently lead to slight variations of the
thin disc properties, but a negligible influence on the SFR.
A variety of predictions were made concerning the number density, age 
and metallicity distributions of stellar subpopulations as a function of $z$
above the galactic plane.
\section*{Acknowledgements}
 
We thank Andrea Borch for providing us with the stellar evolution data with the
PEGASE code.


\end{document}